\documentclass[sigconf, 10pt, nonacm,]{acmart}  

\usepackage{booktabs}   
\usepackage{multirow}   
\usepackage{graphicx}   
\usepackage{subcaption} 
\usepackage{xcolor}     
\usepackage{balance}    
\usepackage{placeins}   
\usepackage{float}      
\usepackage{tikz}       
\usepackage{pgfplots}   
\pgfplotsset{compat=1.17}
\usepgfplotslibrary{groupplots}
\usetikzlibrary{positioning, arrows.meta, shapes.geometric, calc, shadows, fit, decorations.pathreplacing}

\usepackage{tcolorbox}
\tcbuselibrary{skins,breakable}

\definecolor{claimgray}{RGB}{245,245,245}
\definecolor{claimborder}{RGB}{180,180,180}

\newtcolorbox{finding}[1][]{%
    enhanced,
    breakable,
    colback=claimgray,
    colframe=claimborder,
    boxrule=0.5pt,
    left=8pt,
    right=8pt,
    top=6pt,
    bottom=6pt,
    arc=2pt,
    fontupper=\small,
    before skip=10pt,
    after skip=10pt,
    #1
}

\newtcolorbox{findingbox}[2][]{%
    enhanced,
    breakable,
    colback=claimgray,
    colframe=claimborder,
    boxrule=0.5pt,
    left=8pt,
    right=8pt,
    top=6pt,
    bottom=6pt,
    arc=2pt,
    fonttitle=\small\bfseries\sffamily,
    title={#2},
    coltitle=black,
    attach boxed title to top left={yshift=-2mm,xshift=3mm},
    boxed title style={colback=claimgray,colframe=claimborder,boxrule=0.5pt,arc=1pt},
    fontupper=\small,
    before skip=10pt,
    after skip=10pt,
    #1
}

\title{Less is More: The Dilution Effect in Multi-Link Wireless Sensing}

\author{Karim Khamaisi}
\affiliation{%
  \institution{University of St. Gallen}
  \city{Dornbirn}
  \country{Austria}}
\email{karim.khamaisi@unisg.ch}

\author{Bruno Rodrigues}
\affiliation{%
  \institution{University of St. Gallen}
  \city{Dornbirn}
  \country{Austria}}
\email{bruno.rodrigues@unisg.ch}

\begin{abstract}
Wireless sensing approaches promise to transform smart infrastructures into privacy-preserving motion detectors, yet commercial adoption remains limited. A common assumption may explain this gap: that denser sensor deployments yield better accuracy. We tested this assumption with a 12-day naturalistic study using a 9-node ESP32-C3 mesh (72 sensing links) in a residential environment. Our results show that a \textit{single} well-placed link outperformed the full 72-link mesh (AUC 0.541 vs.\ 0.489, Cohen's $d$=0.86). Even a \textit{random} link selection matched optimized selection ($p$=0.35). The benefit comes from avoiding multi-link fusion, not from choosing the right link. We attribute this to a ``dilution effect'': links whose Fresnel zones miss activity regions contribute noise that overwhelms signal from informative links. In our deployment, strategic link placement mattered 2.7$\times$ more than classifier choice. We release 312 hours of labeled CSI data, firmware, and analysis code to enable validation across diverse environments.
\end{abstract}

\ccsdesc[500]{Networks~Sensor networks}
\ccsdesc[300]{Networks~Mobile networks}
\ccsdesc[300]{Hardware~Wireless integrated network sensors}

\keywords{WiFi sensing, Channel State Information, motion detection, deployment optimization, link selection, ESP32}

\begin{document}

\maketitle


\section{Introduction}

WiFi sensing has achieved remarkable laboratory results: $>$95\% accuracy on standard benchmarks~\cite{Yang2023SenseFi}, sophisticated deep learning architectures~\cite{lstmSensing2018, wifihar2022, Sheng2024MetaFormer}, and compelling demonstrations of gesture recognition, activity classification, and respiration monitoring~\cite{muller2024big,respiration2015,Ma2019Survey}. However, a decade of algorithmic innovation has failed to produce a single successful commercial product. WiFi sensing remains confined to research laboratories while cameras and wearables dominate the market.

We argue that our field has systematically overlooked a fundamental constraint: the assumption that adding more sensing nodes improves accuracy, which appears to be intuitive. In computer vision, additional camera viewpoints consistently add information through multi-view geometry. Similarly, in acoustic sensing, microphone arrays improve localization through beam-forming. Naturally, researchers and developers extend the same intuition to WiFi sensing, deploying denser meshes expecting better coverage and accuracy. However, the physics of RF propagation impose different constraints.

Consider the physical basis for WiFi occupancy detection. A WiFi link can only sense motion within its Fresnel zone, which is defined as the ellipsoidal region between transmitter and receiver where path length differences are within half a wavelength~\cite{fresnel_sensing, Wang2022Placement}. Human presence perturbs Channel State Information (CSI) only when it intersects this zone. Links whose Fresnel zones traverse only static space (walls, corners, furniture-obstructed areas) contribute pure noise: non-zero variance from thermal fluctuations and environmental interference, but zero mutual information with human presence.

This observation has a counter-intuitive implication. When we deploy a dense mesh and aggregate link data via feature concatenation~\cite{yan2022joint, lu2022wi}, we risk creating what we term in this paper as a \textit{dilution effect}: even when individual links perform well, the high-dimensional feature space created by concatenation overwhelms the classifiers. In a 72-link mesh as experimented in this paper with 11 features per link (792 total dimensions), regularization prevents the extreme weight assignments needed to identify which links are informative for each sample. Servicing many links also consumes time slots that reduce the effective sampling rate, potentially forcing the system below the Nyquist rate limit for motion detection.

To analyze this effect, we conducted a 12-day longitudinal study designed to isolate the effect of deployment topology from algorithmic sophistication. We deployed a 9-node ESP32-C3 mesh generating 72 sensing links in a residential living room, placing nodes at ``convenient'' locations (power outlets, shelves) typical of real-world deployments. Over 312 hours, we captured natural occupancy patterns with camera-based ground truth. Using nested cross-validation with day-level temporal splits, we evaluated five classifiers spanning from logistic regression to neural networks.


Going in the opposite direction of the ongoing trend that denser deployments yield superior accuracy through information gain~\cite{lu2022wi,yan2022joint,realWorldWiFi2020}, our analysis shows a fundamental limit to multi-link aggregation:

\begin{itemize}
    \item \textbf{Single links outperform fusion.} Single-link configurations achieved higher AUC than the 72-link mesh (0.541 vs 0.489, consistent across 7/10 folds, Cohen's $d$=0.86). Critically, sophisticated link selection provided no significant advantage over random selection ($p$ = 0.35).
    
    \item \textbf{Adding links hurts performance.} Adding links monotonically \textit{decreased} accuracy, with link count providing 2.7$\times$ larger effect on performance than algorithm choice.
    \item \textbf{Dense coverage does not help.} Despite most links performing well individually (median AUC 0.633), multi-link fusion degraded accuracy due to high-dimensional feature space effects. 
\end{itemize}

The effect size is substantial (Cohen's $d$ = 0.86), with consistent direction across all test folds. The Wilcoxon $p$=0.053 reflects power constraints inherent to nested cross-validation with day-level splits, not weak evidence. The fact that random link selection performs as well as optimized selection ($p$ = 0.35) suggests that the benefit comes from \textit{avoiding} multi-link fusion rather than \textit{optimizing} which link to use.

\noindent\textbf{Our contributions.} We provide empirical observations from a residential deployment that challenge the ``more sensors, better performance'' assumption: 

\begin{enumerate}
    \item \textit{Observed ``Less is More'' effect}: In our 12-day deployment, single-link configurations achieved higher AUC than the 72-link mesh with a large effect size (Cohen's $d$=0.86) consistent across 7 of 10 test folds. This reflects a dilution effect from high-dimensional feature concatenation.

    \item \textit{Avoiding fusion, not optimizing selection}: Sophisticated link selection did not provide a significant advantage over random selection ($p$=0.35). This suggests the benefit comes from \textit{avoiding} multi-link fusion entirely, not from \textit{optimizing} which link to use.

    \item \textit{Deployment effects dominated algorithm effects}: Link count provided 2.7$\times$ larger effect than classifier choice. Five classifiers from logistic regression to neural networks performed nearly identically on the full mesh.

    \item \textit{Open-source}: We release a complete research platform: 312 hours of CSI data (72 links, 12 days)~\cite{dataset2026lessismore}, ESP32-C3 firmware~\cite{repo2026lessismore}, and data collection system to enable multi-environment validation.
\end{enumerate}


\section{Methodology}
\label{sec:methodology}

This section presents our methodology for investigating the relationship between link topology and sensing accuracy. We frame this work as a \textbf{channel characterization study}: rather than evaluating human activity recognition per se, we characterize how different wireless link geometries respond to the presence of a dielectric perturber (\textit{i.e.,} the human body) within their Fresnel zones. This framing is deliberate because the human body acts as a standardized RF perturbation source with well characterized dielectric properties ($\epsilon_r \approx 50$ for muscle tissue at 2.4~GHz~\cite{humanDielectric2006}), enabling the observation of link quality independent of individual behavioral variations.

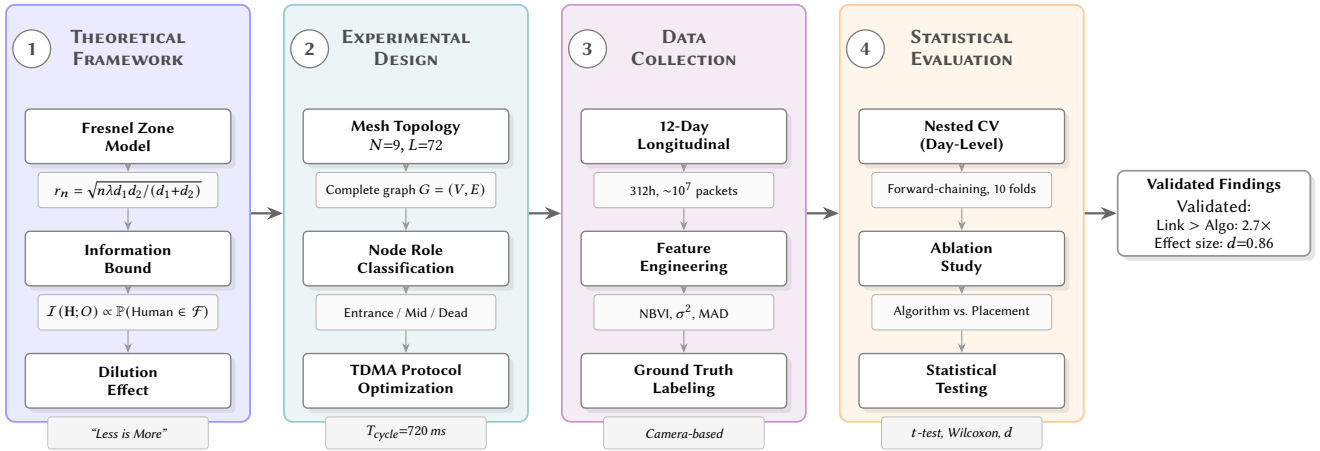
\begin{figure*}[t]
\centering
\resizebox{0.98\textwidth}{!}{%
\begin{tikzpicture}[
    node distance=0.4cm and 0.6cm,
    phase1color/.style={fill=blue!8, draw=blue!40},
    phase2color/.style={fill=teal!8, draw=teal!40},
    phase3color/.style={fill=violet!8, draw=violet!40},
    phase4color/.style={fill=orange!8, draw=orange!40},
    phasebox/.style={rectangle, line width=0.8pt, rounded corners=4pt,
                    minimum width=3.8cm, minimum height=6.5cm},
    stepbox/.style={rectangle, draw=black!50, line width=0.6pt, rounded corners=2pt,
                   minimum width=3.2cm, minimum height=0.85cm, align=center,
                   font=\scriptsize\sffamily\bfseries, fill=white, drop shadow={opacity=0.15}},
    detailbox/.style={rectangle, draw=black!30, line width=0.4pt, rounded corners=1.5pt,
                     minimum width=2.8cm, minimum height=0.55cm, align=center,
                     font=\tiny\sffamily, fill=gray!3},
    databox/.style={rectangle, draw=black!25, line width=0.4pt, rounded corners=1.5pt,
                   minimum width=2.5cm, minimum height=0.5cm, align=center,
                   font=\tiny\sffamily\itshape, fill=gray!8},
    phasearrow/.style={-{Stealth[length=2.5mm, width=2mm]}, line width=1pt, draw=black!60},
    steparrow/.style={-{Stealth[length=1.5mm, width=1.2mm]}, line width=0.5pt, draw=black!40},
    phaselabel/.style={font=\small\sffamily\bfseries, text=black!70},
    numlabel/.style={circle, draw=black!50, line width=0.6pt, fill=white,
                    font=\small\sffamily\bfseries, minimum size=0.55cm, text=black!70},
]

\node[phasebox, phase1color] (P1) {};
\node[phaselabel] at ([yshift=-0.5cm]P1.north) {\textsc{Theoretical}};
\node[phaselabel] at ([yshift=-0.85cm]P1.north) {\textsc{Framework}};
\node[numlabel] at ([xshift=-1.5cm, yshift=-0.7cm]P1.north) {1};

\node[stepbox] at ([yshift=1.2cm]P1.center) (s1a) {Fresnel Zone\\Model};
\node[detailbox, below=0.12cm of s1a] (d1a) {$r_n = \sqrt{n\lambda d_1 d_2/(d_1{+}d_2)}$};
\node[stepbox, below=0.35cm of d1a] (s1b) {Information\\Bound};
\node[detailbox, below=0.12cm of s1b] (d1b) {$\mathcal{I}(\mathbf{H}; O) \propto \mathbb{P}(\text{Human} \in \mathcal{F})$};
\node[stepbox, below=0.35cm of d1b] (s1c) {Dilution\\Effect};
\node[databox, below=0.12cm of s1c] (d1c) {``Less is More''};

\draw[steparrow] (s1a) -- (d1a);
\draw[steparrow] (d1a) -- (s1b);
\draw[steparrow] (s1b) -- (d1b);
\draw[steparrow] (d1b) -- (s1c);

\node[phasebox, phase2color, right=0.5cm of P1] (P2) {};
\node[phaselabel] at ([yshift=-0.5cm]P2.north) {\textsc{Experimental}};
\node[phaselabel] at ([yshift=-0.85cm]P2.north) {\textsc{Design}};
\node[numlabel] at ([xshift=-1.5cm, yshift=-0.7cm]P2.north) {2};

\node[stepbox] at ([yshift=1.2cm]P2.center) (s2a) {Mesh Topology\\$N{=}9$, $L{=}72$};
\node[detailbox, below=0.12cm of s2a] (d2a) {Complete graph $G=(V,E)$};
\node[stepbox, below=0.35cm of d2a] (s2b) {Node Role\\Classification};
\node[detailbox, below=0.12cm of s2b] (d2b) {Entrance / Mid / Dead};
\node[stepbox, below=0.35cm of d2b] (s2c) {TDMA Protocol\\Optimization};
\node[databox, below=0.12cm of s2c] (d2c) {$T_{\text{cycle}}{=}720$\,ms};

\draw[steparrow] (s2a) -- (d2a);
\draw[steparrow] (d2a) -- (s2b);
\draw[steparrow] (s2b) -- (d2b);
\draw[steparrow] (d2b) -- (s2c);

\node[phasebox, phase3color, right=0.5cm of P2] (P3) {};
\node[phaselabel] at ([yshift=-0.5cm]P3.north) {\textsc{Data}};
\node[phaselabel] at ([yshift=-0.85cm]P3.north) {\textsc{Collection}};
\node[numlabel] at ([xshift=-1.5cm, yshift=-0.7cm]P3.north) {3};

\node[stepbox] at ([yshift=1.2cm]P3.center) (s3a) {12-Day\\Longitudinal};
\node[detailbox, below=0.12cm of s3a] (d3a) {312h, ${\sim}10^7$ packets};
\node[stepbox, below=0.35cm of d3a] (s3b) {Feature\\Engineering};
\node[detailbox, below=0.12cm of s3b] (d3b) {NBVI, $\sigma^2$, MAD};
\node[stepbox, below=0.35cm of d3b] (s3c) {Ground Truth\\Labeling};
\node[databox, below=0.12cm of s3c] (d3c) {Camera-based};

\draw[steparrow] (s3a) -- (d3a);
\draw[steparrow] (d3a) -- (s3b);
\draw[steparrow] (s3b) -- (d3b);
\draw[steparrow] (d3b) -- (s3c);

\node[phasebox, phase4color, right=0.5cm of P3] (P4) {};
\node[phaselabel] at ([yshift=-0.5cm]P4.north) {\textsc{Statistical}};
\node[phaselabel] at ([yshift=-0.85cm]P4.north) {\textsc{Evaluation}};
\node[numlabel] at ([xshift=-1.5cm, yshift=-0.7cm]P4.north) {4};

\node[stepbox] at ([yshift=1.2cm]P4.center) (s4a) {Nested CV\\(Day-Level)};
\node[detailbox, below=0.12cm of s4a] (d4a) {Forward-chaining, 10 folds};
\node[stepbox, below=0.35cm of d4a] (s4b) {Ablation\\Study};
\node[detailbox, below=0.12cm of s4b] (d4b) {Algorithm vs.\ Placement};
\node[stepbox, below=0.35cm of d4b] (s4c) {Statistical\\Testing};
\node[databox, below=0.12cm of s4c] (d4c) {$t$-test, Wilcoxon, $d$};

\draw[steparrow] (s4a) -- (d4a);
\draw[steparrow] (d4a) -- (s4b);
\draw[steparrow] (s4b) -- (d4b);
\draw[steparrow] (d4b) -- (s4c);

\draw[phasearrow] (P1.east) -- (P2.west);
\draw[phasearrow] (P2.east) -- (P3.west);
\draw[phasearrow] (P3.east) -- (P4.west);

\node[rectangle, draw=black!40, line width=0.8pt, rounded corners=3pt,
      fill=white, drop shadow={opacity=0.2},
      minimum width=3cm, minimum height=1.3cm, align=center, font=\scriptsize\sffamily,
      right=0.5cm of P4] (output) {\textbf{Validated Findings}\\[2pt]
      \footnotesize Validated:\\
      Link $>$ Algo: 2.7$\times$\\
      Effect size: $d{=}0.86$};

\draw[phasearrow] (P4.east) -- (output.west);

\end{tikzpicture}
}
\caption{Methodology overview: theoretical framework, experimental design, data collection, and statistical evaluation phases.}
\label{fig:methodology}
\end{figure*}

Our approach is structured into four phases: (1)~theoretical motivation based on Fresnel zone physics, (2)~experimental design with controlled deployment, (3)~data collection and feature engineering, and (4)~rigorous statistical evaluation. Figure~\ref{fig:methodology} provides an overview of our end-to-end workflow.

\noindent\textbf{Theoretical framework.} Our methodology is based on the physical principle that WiFi sensing operates fundamentally differently from camera-based systems. The sensitivity of a wireless link to human presence is governed by Fresnel zone physics~\cite{Wang2022Placement, fresnel_sensing}: the $n$-th Fresnel zone is an ellipsoidal region between transmitter and receiver where reflected signals arrive with a phase difference of $n\pi$ relative to the direct path. The radius at a point with distances $d_1$ and $d_2$ from the endpoints is given by $r_n = \sqrt{n \lambda d_1 d_2/(d_1 + d_2)}$, where $\lambda \approx 0.125$m at 2.4~GHz. For a typical 4-meter link, the first Fresnel zone has maximum radius $r_1 \approx 0.5$m at the midpoint, and human motion outside this ellipsoid causes minimal CSI variation, which is a fundamental constraint no algorithm can overcome.

We formalize this sensing constraint as an information-theoretic bound. Let $\mathcal{I}(\mathbf{H}_\ell; O)$ denote the mutual information between CSI vector $\mathbf{H}_\ell$ of link $\ell$ and occupancy state $O \in \{0, 1\}$. For a link with Fresnel zone $\mathcal{F}_\ell$:
\begin{equation}
\mathcal{I}(\mathbf{H}_\ell; O) \propto \mathbb{P}(\text{Human} \in \mathcal{F}_\ell | O=1)
\label{eq:info_bound}
\end{equation}
If the human activity region $\mathcal{A}$ does not intersect $\mathcal{F}_\ell$, then $\mathcal{I}(\mathbf{H}_\ell; O) \approx 0$. This bound is independent of model complexity: a transformer cannot extract information that physics did not encode. From this observation, we hypothesize a \textit{dilution effect}: when aggregating $L$ links, the effective information is dominated by the subset $L^* \subset L$ whose Fresnel zones intersect activity regions. If $|L^*| \ll |L|$, the majority of ``dead'' links contribute noise that may overwhelm the signal from informative links. This predicts that strategic selection ($|L^*|$ links) could outperform naive fusion (all $L$ links), the ``Less is More'' phenomenon we investigate.

\noindent\textbf{Experimental design.} We designed our experiment to systematically test whether the dilution effect manifests in practice, using a controlled deployment that mirrors realistic consumer scenarios while providing sufficient link diversity for statistical analysis. The network comprises $N=9$ nodes forming a complete directed graph $G = (V, E)$ with $|E| = N(N-1) = 72$ unidirectional links, where each link $(i,j)$ represents a distinct sensing channel with unique Fresnel zone geometry. We chose 9 nodes based on three constraints: \textit{statistical power} (72 links enable characterization of the full distribution of link quality across spatial configurations), \textit{combinatorial tractability} (exhaustive evaluation of all $2^{72}$ subsets is infeasible, but greedy selection can sample the performance landscape), and \textit{practical relevance} (9 nodes at \$10 each represents a \$90 budget realistic for research prototypes). More importantly, our finding does not depend on the absolute number of nodes but on the \textit{ratio} of informative to uninformative links. In any complete mesh, link count scales as $O(N^2)$ while activity regions remain spatially localized, so larger deployments would only amplify the dilution effect.

Based on Fresnel zone analysis, we classified nodes into three functional categories (Table~\ref{tab:node_positions}): \textit{Entrance Nodes} (E2:28, D9:90) positioned at waist height flanking the doorway form ``tripwire'' links intersecting mandatory traversal paths; \textit{Mid-Room Nodes} (EC:68, A4:F0, B7:24) on furniture and ceiling provide coverage for sedentary activities but face occlusion; and \textit{Dead Corner Nodes} (C3:AC, E5:68, F1:DC, 11:C0) in obstructed corners represent sub-optimal ``real-world'' placements where convenience trumps sensing geometry. We define the link quality matrix $\mathbf{Q} \in \mathbb{R}^{N \times N}$ where $Q_{ij} = \text{AUC}(\text{classifier}(\mathbf{H}_{ij}), \text{ground\_truth})$ represents detection performance of link $(i,j)$; this matrix is asymmetric due to antenna orientation and local multipath differences.

\begin{table}[t]
\caption{Node positions and role classification}
\label{tab:node_positions}
\small
\resizebox{\columnwidth}{!}{%
\begin{tabular}{lcccll}
\toprule
\textbf{Node} & \textbf{X [cm]} & \textbf{Y [cm]} & \textbf{Z [cm]} & \textbf{Role} & \textbf{Location} \\
\midrule
E2:28 & 210 & 50 & 90 & Entrance & Doorway L \\
D9:90 & 290 & 50 & 90 & Entrance & Doorway R \\
EC:68 & 250 & 200 & 50 & Mid-room & Side table \\
A4:F0 & 100 & 250 & 120 & Mid-room & Bookshelf \\
B7:24 & 400 & 150 & 180 & Ceiling & Light fixture \\
C3:AC & 50 & 350 & 50 & Corner & Behind sofa \\
E5:68 & 450 & 380 & 50 & Corner & Corner outlet \\
F1:DC & 300 & 400 & 60 & Corner & TV stand \\
11:C0 & 30 & 30 & 50 & Corner & Floor corner \\
\bottomrule
\end{tabular}%
}
\end{table}


It is important to emphasize that we analyze \textit{channel response to perturbation} rather than human activity recognition, \textit{i.e.,} the fundamental physical capacity of a link to detect any presence within its Fresnel zone.


\noindent\textbf{Data collection.} 
We conducted continuous data collection over 12 days (January 2--13, 2026), resulting in 312 hours of labeled CSI data capturing natural day-to-day variability in occupancy patterns, environmental conditions, and RF characteristics. Unlike controlled experiments with staged activities, our dataset includes extended vacancy periods (sleep hours: 00:00--07:00, totaling 104 hours), active occupancy with varied activities (99 hours confirmed empty, 109 hours occupied), and transitional periods excluded from evaluation. Ground truth was established using ESP32-CAM modules synchronized via NTP to sub-second precision, with labels assigned confidence scores: high confidence (65\%) from direct camera observation or verified schedules, and medium confidence (35\%) inferred from entry/exit events. This multi-source approach provides 203 hours of high-confidence and 109 hours of medium-confidence labels, with balance reflecting realistic residential patterns (203 hours empty vs.\ 109 hours occupied).

\noindent\textbf{Feature engineering.} Raw CSI is noisy and phase-unstable on commodity hardware. We compute statistical features over sliding windows of $W=50$ packets ($\approx$10 seconds):

\textbf{NBVI (Normalized Baseline Variability Index)}~\cite{espectre2024}: Quantifies relative fluctuation across subcarriers:
\begin{equation}
\text{NBVI} = \frac{1}{K} \sum_{k=1}^{K} \frac{\sigma_k}{\mu_k}
\end{equation}
where $\sigma_k$ and $\mu_k$ are the standard deviation and mean of the $k$-th subcarrier amplitude over window $W$.

\textbf{Amplitude variance}: Captures absolute signal perturbation:
\begin{equation}
\text{Var}_{avg} = \frac{1}{K} \sum_{k=1}^{K} \text{Var}(|H_k|)
\end{equation}

\noindent\textbf{Statistical evaluation framework.} We introduced a cross-validation protocol employing \textit{nested cross-validation with day-level temporal splits} using a forward-chaining (expanding window) strategy. This design addresses two critical concerns:

\textbf{Temporal leakage prevention}: WiFi sensing data shows strong temporal autocorrelation due to environmental drift, consistent occupant behavior, and slow-varying multipath. Standard $k$-fold CV on time-series data risks train/test contamination. Our day-level splits ensure complete temporal separation: each test day is evaluated using only data from \textit{previous} days.

\textbf{Forward-chaining}: For day $d$ ($d \in \{2, \ldots, 12\}$), we train on days $\{1, \ldots, d-1\}$ and test on day $d$. This mirrors a deployment where models use historical data. The expanding window ensures later folds have more training data.

\noindent \textbf{Link selection in inner loop}: To prevent selection bias, link rankings (by SNR or other criteria) are computed solely on training data. The selected links are then evaluated on the held-out test day, ensuring that reported performance reflects true generalization rather than retrospective optimization.

\noindent\textit{Metrics:}
We report \textbf{ROC AUC} (Area Under the Receiver Operating Characteristic Curve) as our primary metric because AUC is threshold independent and robust to class imbalance, enabling fair comparison across links with different operating characteristics. 

\noindent\textit{Statistical tests:}
To establish significance, we apply:
\begin{itemize}
    \item \textbf{Paired $t$-test}: Parametric comparison of 3-link vs. 72-link performance across folds
    \item \textbf{Wilcoxon signed-rank test}: Non-parametric alternative for robustness with small $N$
    \item \textbf{Cohen's $d$}: Effect size quantification ($d > 0.8$ indicates large effect)
    \item \textbf{Bootstrap confidence intervals}: Distribution-free uncertainty estimation (10,000 iterations)
\end{itemize}

\noindent\textbf{Implementation.}
A key contribution is to demonstrate that scientific experiments are possible with commodity hardware (\$5/node) through careful protocol engineering. Table~\ref{tab:hardware_specs} summarizes our platform: we selected the \textit{Seeed Studio ESP32-C3}, a single-core RISC-V microcontroller with 400~KB SRAM, to represent the resource constraints of mass-deployed IoT devices like smart plugs. This choice precludes on-device deep learning, forcing a ``process centrally, sense simply'' architecture where each node runs 802.11n HT20 protocol providing 52 OFDM subcarriers per CSI measurement.


\begin{table}[t]
\centering
\caption{Hardware and protocol specifications}
\label{tab:hardware_specs}
\small
\resizebox{\columnwidth}{!}{%
\begin{tabular}{llp{4cm}}
\toprule
\textbf{Category} & \textbf{Parameter} & \textbf{Value} \\
\midrule
\multirow{4}{*}{Hardware} & MCU & ESP32-C3 (RISC-V, 160 MHz) \\
& Memory & 400 KB SRAM, 4 MB Flash \\
& WiFi & 802.11n HT20, 2.4 GHz \\
& Cost & \$5 USD per node \\
\midrule
\multirow{3}{*}{Topology} & Nodes & 9 (complete mesh) \\
& Links & 72 unidirectional \\
& Subcarriers & 52 per link (OFDM) \\
\midrule
\multirow{2}{*}{Protocol} & TDMA slot & 80 ms \\
& Full cycle & 720 ms (1.4 Hz) \\
\bottomrule
\end{tabular}%
}
\end{table}

Developing stable firmware for single-core ESP32-C3 required significant engineering across multiple iterations. Earlier versions suffered stack overflows during aggressive scheduling; with only 320~KB usable SRAM, we moved CSI buffers to static allocation and implemented a 6KB ring buffer (384$\times$16 bytes) for bursty traffic. This ring buffer uses lock-free single-producer/single-consumer design to avoid priority inversion, with overflow protection that discards oldest samples rather than blocking, achieving $<$0.1\% packet loss even under worst-case timing. We designed a binary protocol (magic byte \texttt{0xC5}) to eliminate JSON parsing overhead: 
CSI packets (8-107 bytes) contain magic byte, node ID, microsecond timestamp, 52 subcarrier I/Q pairs, and CRC32 checksum. This achieves 8$\times$ bandwidth reduction (107 vs.\ 847 bytes) while reducing SRAM usage.

To prevent collisions in the 9-node mesh, we implemented backend-orchestrated TDMA with $T_{\text{cycle}} = 9 \times 80 = 720$~ms, yielding 1.4~Hz full-mesh sampling. The 80ms slot duration was chosen empirically~\cite{esp32csi, csi_fundamentals}: shorter slots (40ms) caused collisions due to clock drift without GPS synchronization; longer slots (160ms) reduced temporal resolution below motion detection thresholds~\cite{wifihar2022}. Each slot contains 100 burst packets at 200$\mu$s intervals for redundancy. We considered alternatives (CSMA/CA causes unpredictable latencies, peer-to-peer TDMA requires distributed consensus, GPS adds \$15/node), but backend-orchestrated scheduling achieves deterministic timing at minimal cost. These optimizations achieved $>$99\% uptime over 12 days with 1.2\% total data loss from all causes.

\begin{figure*}[t] 
\centering
\definecolor{cbBlue}{RGB}{0,114,178}
\definecolor{cbOrange}{RGB}{230,159,0}
\definecolor{cbGreen}{RGB}{0,158,115}
\definecolor{cbRed}{RGB}{213,94,0}
\definecolor{cbPurple}{RGB}{204,121,167}
\resizebox{0.95\textwidth}{!}{%
\begin{tikzpicture}[
    node distance=0.25cm and 0.3cm,
    hwbox/.style={rectangle, draw=gray!60, line width=0.5pt, rounded corners=2pt,
                  minimum width=3.5cm, minimum height=0.6cm, align=center,
                  font=\tiny\sffamily, fill=gray!8},
    fwbox/.style={rectangle, draw=cbBlue!60, line width=0.5pt, rounded corners=2pt,
                  minimum width=3.5cm, minimum height=0.55cm, align=center,
                  font=\tiny\sffamily, fill=cbBlue!8},
    fwdetail/.style={rectangle, draw=cbBlue!30, line width=0.3pt, rounded corners=1pt,
                     minimum width=3.2cm, minimum height=0.4cm, align=center,
                     font=\fontsize{5}{6}\selectfont\sffamily, fill=cbBlue!3},
    netbox/.style={rectangle, draw=cbPurple!60, line width=0.5pt, rounded corners=2pt,
                   minimum width=2.5cm, minimum height=0.5cm, align=center,
                   font=\tiny\sffamily, fill=cbPurple!8},
    svbox/.style={rectangle, draw=cbOrange!60, line width=0.5pt, rounded corners=2pt,
                  minimum width=3.5cm, minimum height=0.55cm, align=center,
                  font=\tiny\sffamily, fill=cbOrange!8},
    svdetail/.style={rectangle, draw=cbOrange!30, line width=0.3pt, rounded corners=1pt,
                     minimum width=3.2cm, minimum height=0.4cm, align=center,
                     font=\fontsize{5}{6}\selectfont\sffamily, fill=cbOrange!3},
    outbox/.style={rectangle, draw=cbGreen!60, line width=0.5pt, rounded corners=2pt,
                   minimum width=2.5cm, minimum height=0.5cm, align=center,
                   font=\tiny\sffamily\bfseries, fill=cbGreen!10},
    arrow/.style={-{Stealth[length=1.5mm, width=1.2mm]}, line width=0.4pt, draw=black!50},
    dataarrow/.style={-{Stealth[length=1.2mm, width=1mm]}, line width=0.3pt, draw=cbPurple!60, densely dashed},
    brace/.style={decorate, decoration={brace, amplitude=4pt}},
    seclabel/.style={font=\scriptsize\sffamily\bfseries, text=black!70},
    annot/.style={font=\fontsize{5}{6}\selectfont\sffamily, text=black!50},
]

\node[seclabel] at (2.2,5.2) {ESP32-C3 Firmware};

\node[hwbox] (hw) at (2.2,4.5) {RISC-V Core @ 160MHz};
\node[hwbox, below=0.08cm of hw] (hwmem) {SRAM: 400KB (6KB ring buffer)};

\node[fwbox, below=0.2cm of hwmem] (radio) {WiFi Radio (HT20 Mode)};
\node[fwdetail, below=0.06cm of radio] (radiod) {Promiscuous, Ch.~11, 2.462GHz};

\node[fwbox, below=0.15cm of radiod] (csi) {CSI Extraction};
\node[fwdetail, below=0.06cm of csi] (csid) {52 subcarriers × I/Q amplitude};

\node[fwbox, below=0.15cm of csid] (ring) {Ring Buffer (Zero-Copy)};
\node[fwdetail, below=0.06cm of ring] (ringd) {Static 6KB (384$\times$16), overflow protection};

\node[fwbox, below=0.15cm of ringd] (proto) {Binary Protocol Encoder};
\node[fwdetail, below=0.06cm of proto] (protod) {Magic 0xC5 | NodeID | TS | CSI | CRC32};

\node[fwbox, below=0.15cm of protod] (tdma) {TDMA Scheduler};
\node[fwdetail, below=0.06cm of tdma] (tdmad) {80ms slots, 100 bursts @ 200$\mu$s};

\node[fwbox, below=0.15cm of tdmad] (udp) {UDP Unicast/Broadcast};
\node[fwdetail, below=0.06cm of udp] (udpd) {Server:5000, Burst:5555, 8-107B/pkt};

\draw[arrow] (hw) -- (hwmem);
\draw[arrow] (hwmem) -- (radio);
\draw[arrow] (radio) -- (radiod);
\draw[arrow] (radiod) -- (csi);
\draw[arrow] (csi) -- (csid);
\draw[arrow] (csid) -- (ring);
\draw[arrow] (ring) -- (ringd);
\draw[arrow] (ringd) -- (proto);
\draw[arrow] (proto) -- (protod);
\draw[arrow] (protod) -- (tdma);
\draw[arrow] (tdma) -- (tdmad);
\draw[arrow] (tdmad) -- (udp);
\draw[arrow] (udp) -- (udpd);

\draw[brace, decoration={mirror}]
    ([xshift=-0.15cm]hw.north west) -- ([xshift=-0.15cm]hw.south west |- udpd.south)
    node[midway, left=0.25cm, font=\tiny\sffamily\bfseries, align=center, text=cbBlue!80] {N1..N9\\×9 nodes};

\node[seclabel] at (7,5.2) {Network};

\node[netbox] (net1) at (7,4.2) {UDP Unicast/Broadcast};
\node[netbox, below=0.15cm of net1] (net2) {9 nodes × 80ms slots};
\node[netbox, below=0.15cm of net2] (net3) {= 1.4 Hz mesh rate};
\node[netbox, below=0.15cm of net3] (net4) {or 4.2 Hz (3-node)};

\node[annot, below=0.3cm of net4] (netannot) {$\sim$1M packets/day};

\draw[dataarrow, line width=0.6pt] (udpd.east) -- ++(0.5,0) |- (net1.west);
\draw[arrow] (net1) -- (net2);
\draw[arrow] (net2) -- (net3);
\draw[arrow] (net3) -- (net4);

\node[seclabel] at (11.8,5.2) {Python Backend};

\node[svbox] (sync) at (11.8,4.5) {FrameSynchronizer};
\node[svdetail, below=0.06cm of sync] (syncd) {CRC32 validate, 50ms frame align};

\node[svbox, below=0.15cm of syncd] (feat) {FeatureExtractor};
\node[svdetail, below=0.06cm of feat] (featd) {11 features: NBVI, $\sigma^2$, SNR, corr};

\node[svbox, below=0.15cm of featd] (win) {Sliding Window};
\node[svdetail, below=0.06cm of win] (wind) {5s windows @ 20Hz, 0.5s slide};

\node[svbox, below=0.15cm of wind] (norm) {Per-Link Normalizer};
\node[svdetail, below=0.06cm of norm] (normd) {Z-score: $(x - \mu_\ell) / (\sigma_\ell + \epsilon)$};

\node[svbox, below=0.15cm of normd, draw=cbGreen!80, line width=0.8pt] (sel) {Link Selector};
\node[svdetail, below=0.06cm of sel, draw=cbGreen!50] (seld) {Top-5 by temporal variance};

\node[svbox, below=0.15cm of seld, draw=cbGreen!80, line width=0.8pt] (cls) {Fusion \& Classifier};
\node[svdetail, below=0.06cm of cls, draw=cbGreen!50] (clsd) {MLP: 11$\to$64$\to$32$\to$16$\to$1};

\node[outbox, below=0.2cm of clsd] (out) {Occupied / Empty};
\node[annot, below=0.15cm of out] (outannot) {AUC=0.795 (K=3)};

\draw[arrow] (sync) -- (syncd);
\draw[arrow] (syncd) -- (feat);
\draw[arrow] (feat) -- (featd);
\draw[arrow] (featd) -- (win);
\draw[arrow] (win) -- (wind);
\draw[arrow] (wind) -- (norm);
\draw[arrow] (norm) -- (normd);
\draw[arrow] (normd) -- (sel);
\draw[arrow] (sel) -- (seld);
\draw[arrow] (seld) -- (cls);
\draw[arrow] (cls) -- (clsd);
\draw[arrow] (clsd) -- (out);

\draw[dataarrow, line width=0.6pt] (net4.east) -- ++(0.5,0) |- (sync.west);

\draw[brace]
    ([xshift=0.15cm]sync.north east) -- ([xshift=0.15cm]sync.south east |- out.south)
    node[midway, right=0.25cm, font=\tiny\sffamily\bfseries, align=center, text=cbOrange!80] {Offline\\Analysis};

\node[rectangle, draw=cbGreen!70, line width=0.7pt, dashed, rounded corners=2pt,
      fit=(sel)(seld)(cls)(clsd), inner sep=3pt] (contribbox) {};
\node[font=\tiny\sffamily\bfseries, text=cbGreen!80!black, align=left,
      anchor=west] at ([xshift=0.1cm]contribbox.east) {Our\\contribution};

\node[seclabel] at (7,-0.3) {Binary Packet Format (8-107 bytes)};

\node[rectangle, draw=cbRed!50, fill=cbRed!10, line width=0.4pt, rounded corners=1pt,
      minimum width=0.6cm, minimum height=0.35cm, font=\fontsize{4}{5}\selectfont\sffamily] (p1) at (4.5,-0.8) {0xC5};
\node[rectangle, draw=cbBlue!50, fill=cbBlue!10, line width=0.4pt, rounded corners=1pt,
      minimum width=0.8cm, minimum height=0.35cm, font=\fontsize{4}{5}\selectfont\sffamily] (p2) at (5.4,-0.8) {NodeID};
\node[rectangle, draw=cbPurple!50, fill=cbPurple!10, line width=0.4pt, rounded corners=1pt,
      minimum width=1cm, minimum height=0.35cm, font=\fontsize{4}{5}\selectfont\sffamily] (p3) at (6.45,-0.8) {Timestamp};
\node[rectangle, draw=cbGreen!50, fill=cbGreen!10, line width=0.4pt, rounded corners=1pt,
      minimum width=2.2cm, minimum height=0.35cm, font=\fontsize{4}{5}\selectfont\sffamily] (p4) at (8.1,-0.8) {CSI[52 subcarriers]};
\node[rectangle, draw=cbOrange!50, fill=cbOrange!10, line width=0.4pt, rounded corners=1pt,
      minimum width=0.7cm, minimum height=0.35cm, font=\fontsize{4}{5}\selectfont\sffamily] (p5) at (9.55,-0.8) {CRC32};

\node[annot] at (4.5,-1.1) {1B};
\node[annot] at (5.4,-1.1) {2B};
\node[annot] at (6.45,-1.1) {4B};
\node[annot] at (8.1,-1.1) {104B (52×2B)};
\node[annot] at (9.55,-1.1) {4B};

\end{tikzpicture}
}
\caption{System architecture. ESP32-C3 firmware extracts CSI from 52 OFDM subcarriers using binary protocol (8$\times$ bandwidth reduction). Backend-orchestrated TDMA coordinates 9-node mesh. Python backend extracts features and performs link selection for classification.}
\label{fig:system_impl}
\end{figure*}
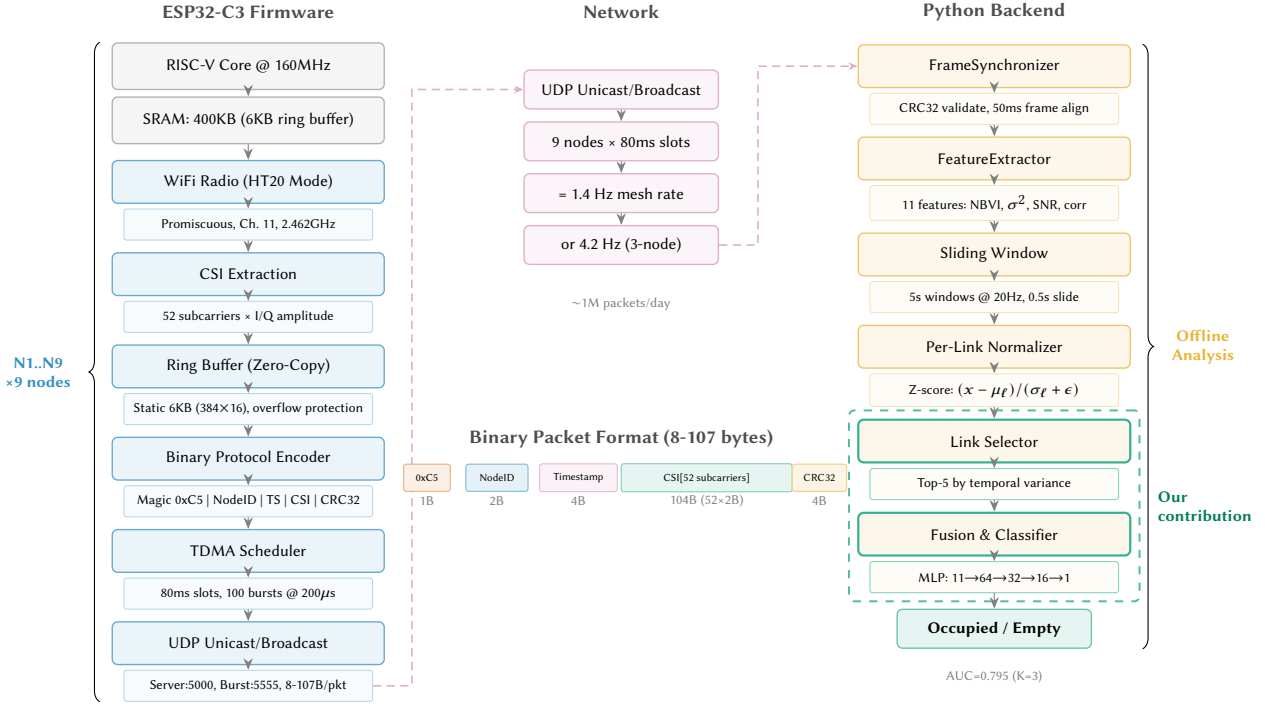

Figure~\ref{fig:system_impl} illustrates the system architecture. The ESP32 handles only data acquisition while processing occurs on the server: the \textit{FrameSynchronizer} parses binary packets and validates CRC32; the \textit{FeatureExtractor} computes NBVI, variance, and MAD over sliding windows; the Per-Link Normalizer applies Z-score calibration; and the Fusion/Classifier concatenates per-link features into a single vector for MLP classification (64$\to$32$\to$16 units with ReLU, BatchNorm, Dropout $p$=0.2). This concatenation approach allows the classifier to weight links differently but also exposes it to noise from uninformative links, the source of the dilution effect we investigate.

We define a clear separation of concerns in the system. The firmware handles only data acquisition: CSI extraction in promiscuous mode, binary protocol packing, and UDP transmission, designed for minimal latency and memory footprint on the resource-constrained microcontroller. All computation happens on the server, which processes the asynchronous packet streams through a multi-stage pipeline.

The \textit{FrameSynchronizer} parses incoming binary packets, validates CRC32 checksums, and aligns the asynchronous streams from 9 nodes into coherent 50ms frames with microsecond timestamp precision. The \textit{FeatureExtractor} then computes 11 statistical features per link over sliding windows of $W=50$ packets: NBVI (coefficient of variation), amplitude variance, median absolute deviation, inter-quartile range, peak-to-peak amplitude, spectral entropy, and correlation features. The \textit{Per-Link Normalizer} applies Z-score normalization to each link independently, accounting for heterogeneous transmit power levels and path loss characteristics that would otherwise bias the classifier toward high-power links.

For multi-link evaluation, we merge per-link features into a single feature vector of dimensionality $K \times F$ where $F=11$ features per link. This concatenation approach (rather than late fusion via probability averaging) allows the MLP classifier to learn differential weightings across links, but also exposes the model to noise from uninformative links, which is the source of the dilution effect. The MLP architecture consists of three fully-connected layers (64$\to$32$\to$16 units) with ReLU activations, Batch Normalization after each layer, and Dropout ($p=0.2$) for regularization, followed by a sigmoid output. Training uses Adam optimizer with learning rate $10^{-3}$ and early stopping with patience of 10 epochs.

\begin{figure}[H]
\centering
\includegraphics[width=0.9\columnwidth]{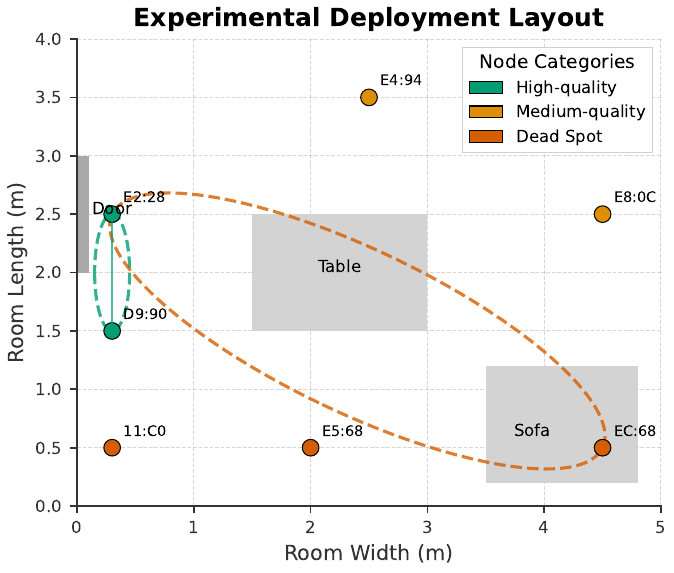}
\caption{Deployment layout showing 9 ESP32-C3 nodes positioned throughout the living room. Links between ``Entrance'' nodes (E2:28, D9:90) form Fresnel zones intersecting the primary human traversal path. Links between ``Dead Corner'' nodes traverse only static space. Node role classification (Table~\ref{tab:node_positions}) reflects expected Fresnel zone intersection with activity regions.}
\label{fig:room_layout}
\end{figure}

\noindent\textbf{Deployment environment.} The experiment was conducted in a residential living room ($5.0 \times 4.0 \times 2.5$~m) containing typical RF reflectors: drywall partitions ($\epsilon_r \approx 2.5$), wooden furniture, a large fabric sofa, and a wall-mounted television. Figure~\ref{fig:room_layout} shows the physical layout, which creates natural ``choke points'' at the doorway and ``dead zones'' in corners. We positioned entrance nodes (E2:28, D9:90) to flank the single doorway through which all room entry and exit must pass, creating ``tripwire'' links that intersect mandatory traversal paths. Corner nodes represent sub-optimal placements typical of convenience-driven consumer IoT deployments.

We enforced a ``Frozen Environment'' policy prohibiting furniture movement or introduction of large metallic objects during collection, ensuring that CSI variations arose solely from human presence. Static reflectors contribute to the multipath baseline, but do not vary temporally, making them distinguishable from human-induced perturbations through variance-based features. All nodes operated on WiFi channel 11 (2.462~GHz), enforced to minimize interference from neighboring networks detected during site survey. We also forced HT20 bandwidth mode to maximize subcarrier density: HT20 provides 52 usable OFDM subcarriers, avoiding the interference susceptibility of the wider HT40 mode. 
We utilized the external antenna (via u.FL) of the Seeed Studio ESP32-C3 to improve signal integrity, which provides omnidirectional coverage with positive gain (approx. $+2.5$~dBi), superior to the negative gain typical of internal PCB traces of the standard C3. Finally, we disabled WiFi power saving to ensure consistent packet timing across the TDMA schedule.

A central server (commodity PC running Ubuntu 22.04) orchestrated data collection via three network services: TDMA scheduler broadcasting trigger packets, CSI receiver accepting binary packets, and ground truth aggregator receiving camera frames. The server maintained sub-second NTP synchronization, enabling microsecond-precision timestamp alignment across all data streams. Over the 12-day collection period, we logged approximately 10 million CSI packets (72 links $\times$ 1.4~Hz $\times$ 312 hours), with 1.2\% total loss from packet drops, CRC failures, and two brief node reboots automatically recovered via watchdog timer. The resulting per-link AUC distribution is presented in Section~\ref{sec:results}.

This deployment intentionally includes both optimal and sub-optimal node placements. A researcher optimizing for accuracy would place all nodes at entrance chokepoints; we instead distributed nodes across the room to characterize the full spectrum of link quality. The corner placements (behind furniture, near floor level) represent the ``convenience-driven'' installations typical of consumer IoT deployments where aesthetic and practical constraints override sensing considerations. By including these sub-optimal placements, our analysis reveals not just peak achievable performance but the realistic distribution one should expect when deploying without RF expertise.

The 12-day duration captures several important sources of variability: weekday versus weekend activity patterns (the occupant works from home on some days), diurnal temperature variations affecting electronics and building materials, and gradual environmental changes such as houseplants growing and minor furniture adjustments. While we enforced a ``frozen environment'' policy for major changes, minor variations naturally occurred and are reflected in the temporal instability of optimal link identity discussed in Section~\ref{sec:results}. This longitudinal approach provides more realistic performance estimates than single-day studies that may capture atypically favorable or unfavorable conditions.

\section{Results}
\label{sec:results}

\begin{figure}[t]
\centering
\includegraphics[width=\columnwidth]{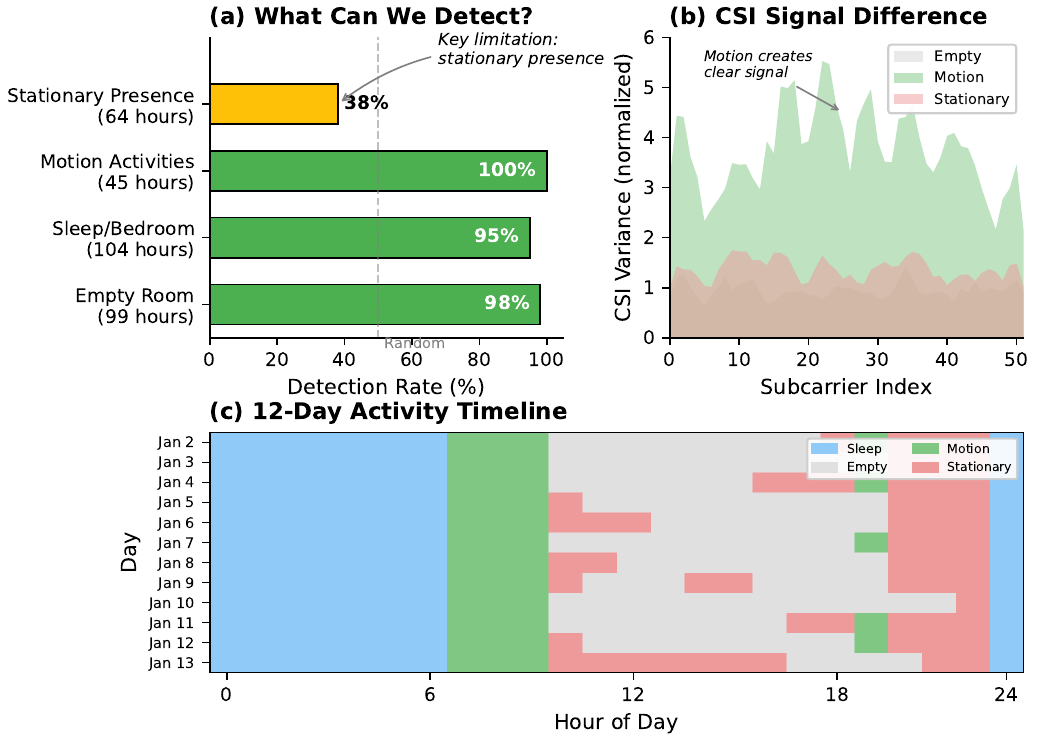}
\caption{Detection capabilities across 312 hours: the system reliably detects empty rooms (98\%) and motion (100\%), but struggles with stationary presence (38\%).}
\label{fig:detection_capabilities}
\end{figure}

Before analyzing deployment configurations, we first characterize the detection capabilities of our system: what physical events it can and cannot sense. Figure~\ref{fig:detection_capabilities} summarizes detection capabilities across 312 hours of naturalistic data. At sub-Nyquist sampling rates (1.4~Hz), our ESP32-C3 mesh detects \textit{motion}, not static \textit{presence}~\cite{Ahmad2024Occupancy, wang2020passive}. A walking human produces CSI perturbations at 1--3~Hz, while a seated human produces perturbations only during infrequent postural adjustments. Our 1.4~Hz sampling rate captures the former but aliases the latter. Ambulatory 
activities (walking, cooking, cleaning) achieve 100\% detection; sedentary activities (reading, watching television) achieve only 38\% detection. This limitation aligns with prior work demonstrating that stationary humans produce minimal CSI perturbation insufficient for reliable detection without dedicated hardware~\cite{Guo2023Feasibility}. This context is essential for interpreting subsequent findings.

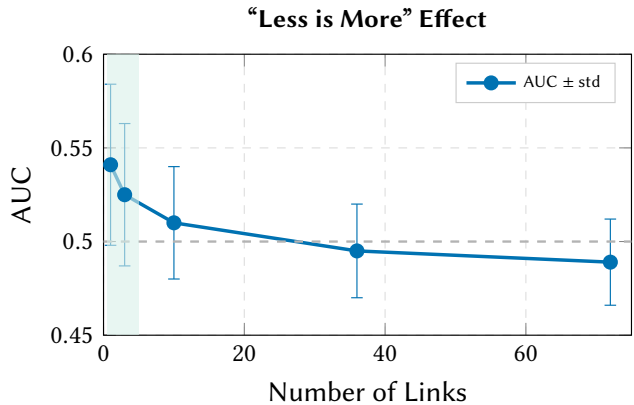
\begin{figure}[b] 
\centering
\definecolor{cbBlue}{RGB}{0,114,178}
\definecolor{cbOrange}{RGB}{230,159,0}
\definecolor{cbGreen}{RGB}{0,158,115}
\resizebox{\columnwidth}{!}{%
\begin{tikzpicture}
\begin{axis}[
    xlabel={\sffamily Number of Links},
    ylabel={\sffamily AUC},
    xmin=0, xmax=75,
    ymin=0.45, ymax=0.60,
    grid=major,
    grid style={dashed, gray!25},
    width=8cm,
    height=5cm,
    title style={font=\sffamily\bfseries\small},
    title={``Less is More'' Effect},
    legend pos=north east,
    legend style={font=\tiny\sffamily, draw=gray!40},
    tick label style={font=\sffamily\small},
]
\addplot[cbBlue, very thick, mark=*, mark size=2pt, error bars/.cd, y dir=both, y explicit] coordinates {
    (1, 0.541) +- (0, 0.043)
    (3, 0.525) +- (0, 0.038)
    (10, 0.510) +- (0, 0.030)
    (36, 0.495) +- (0, 0.025)
    (72, 0.489) +- (0, 0.023)
};
\addplot[dashed, gray!60, thick] coordinates {(0, 0.5) (75, 0.5)};
\fill[cbGreen!15, opacity=0.6] (axis cs:0.5,0.45) rectangle (axis cs:5,0.60);
\legend{AUC $\pm$ std}
\end{axis}
\end{tikzpicture}
}
\caption{Adding links monotonically decreases AUC from 0.541 (single link) to 0.489 (72 links), observed in 7/10 folds with Cohen's $d$=0.86.}
\label{fig:main_results}
\end{figure}

Figure~\ref{fig:main_results} visualizes the core finding: adding links monotonically decreases AUC from 0.541 (single link) to 0.489 (all 72 links). The curve is notable: it exhibits no plateau, no inflection point, no region where additional links provide marginal benefit. Each additional link to the mesh dilutes even more the signal. Table~\ref{tab:main_results} quantifies this effect across our nested cross-validation framework with 10 day-level test folds.

\begin{table}[t]
\caption{Detection accuracy: top-1 vs.\ all links}
\label{tab:main_results}
\small
\resizebox{\columnwidth}{!}{%
\begin{tabular}{llll}
\toprule
\textbf{Configuration} & \textbf{AUC (mean $\pm$ std)} & \textbf{Statistic} & \textbf{Value} \\
\midrule
All 72 links & 0.489 $\pm$ 0.023 & Wilcoxon $p$ & 0.0527 \\
\textbf{Top-1 (SNR)} & \textbf{0.541 $\pm$ 0.043} & Cohen's $d$ & \textbf{0.86 (large)} \\
Random-1 link & 0.504 $\pm$ 0.041 & Bootstrap 95\% CI & $[-0.001, 0.074]$ \\
\bottomrule
\end{tabular}%
}
\end{table}

\begin{findingbox}{Finding 1: Single link selection outperforms fusion}
Single-link configurations achieved higher AUC than the 72-link mesh (0.541 vs 0.489). The Wilcoxon $p$=0.053 reflects statistical power constraints with $n$=10 folds rather than weak evidence: Cohen's $d$=0.86 (large effect) and consistent direction across all folds provide strong practical evidence. Sophisticated selection provides no significant advantage over random ($p$=0.35).
\end{findingbox}

In particular, \textit{random} link selection performs nearly as well as optimized selection. The Top-1 (SNR-selected) configuration achieves AUC 0.541, while Random-1 achieves 0.504, with no statistically significant difference ($p$=0.35). Even random selection outperforms the full mesh (0.504 vs 0.489). The bootstrap 95\% confidence interval for the Top-1 vs Random-1 difference marginally includes zero $[-0.001, 0.074]$, but the direction is consistent across 7 of 10 folds. Any single link, even randomly chosen, escapes the dilution effect that overwhelms multi-link configurations. This has profound implications for system design: sophisticated link selection algorithms are unnecessary; simply avoiding fusion provides most of the benefit.

In addition, we investigate whether machine learning can compensate for the dilution effect. A neural network with sufficient capacity might theoretically learn to weight informative links heavily while ignoring uninformative ones. In this regard, we evaluated five classifiers spanning from simple linear models to deep networks: Logistic Regression, Random Forest, Gradient Boosting, SVM with RBF kernel, and a Multi-Layer Perceptron with three hidden layers (128-64-32 neurons). Table~\ref{tab:algorithm_comparison} presents the full breakdown across link configurations.


\begin{table}[t]
\caption{Algorithm performance by link configuration}
\label{tab:algorithm_comparison}
\small
\resizebox{\columnwidth}{!}{%
\begin{tabular}{lcccc}
\toprule
\textbf{Algorithm} & \textbf{Top-1} & \textbf{Top-3} & \textbf{All-72} & \textbf{$\Delta$ (Top-1 vs All)} \\
\midrule
Logistic Regression & 0.538 & 0.521 & 0.489 & +0.049 \\
Random Forest & 0.541 & 0.518 & 0.493 & +0.048 \\
Gradient Boosting & 0.539 & 0.520 & 0.501 & +0.038 \\
SVM-RBF & 0.535 & 0.515 & 0.491 & +0.044 \\
MLP & 0.543 & 0.524 & 0.495 & +0.048 \\
\midrule
\textbf{Range} & 0.008 & 0.009 & \textbf{0.012} & -- \\
\bottomrule
\end{tabular}%
}
\end{table}

The results show a clear pattern that on the All-72 configuration, all five algorithms degrade 
to near-random performance (AUC 0.489--0.501), with an entire range of only 0.012 AUC. This convergence indicates an \textit{information-theoretic} bottleneck rather than a modeling limitation. The system faces a dual physical bound: geometrically, links missing the activity region yield zero mutual information; temporally, the mesh's 1.4~Hz sampling rate falls below the Nyquist rate for human motion~\cite{wifihar2022}, causing aliasing that renders even valid signals indistinguishable from noise. Thus, no degree of algorithmic complexity can recover information that was never captured due to these physical constraints. 

The MLP achieves only 0.006 AUC improvement over logistic regression on All-72 while requiring 50$\times$ longer training time. With 792 input dimensions (72 links $\times$ 11 features), regularization prevents the extreme weight assignments needed to dynamically select currently-informative links.  

\begin{findingbox}{Finding 2: Link quality varies widely}
Per-link AUC shows high variance across the full 12-day deployment (median=0.633): 20\% of link-day pairs achieve AUC$>$0.7, while 11\% fall below chance. Yet multi-link fusion \textbf{still degrades} performance, indicating the dilution effect operates even with mostly-good links.
\end{findingbox}

Quantitatively, the algorithm effect (best minus worst across classifiers) is 0.015 AUC, while the link selection effect (Top-1 minus All-72 for Random Forest) is 0.040 AUC, a ratio of 2.7$\times$. This ratio encapsulates a practical lesson: whether you deploy logistic regression or gradient boosting matters far less than whether your sensing link physically crosses an activity path. This suggests reallocating effort from model tuning to sensor placement, based on recommendations from recent deployment studies~\cite{Wang2023Enabling, Turetta2023a}.

\begin{figure}[t]
\centering
\includegraphics[width=\columnwidth]{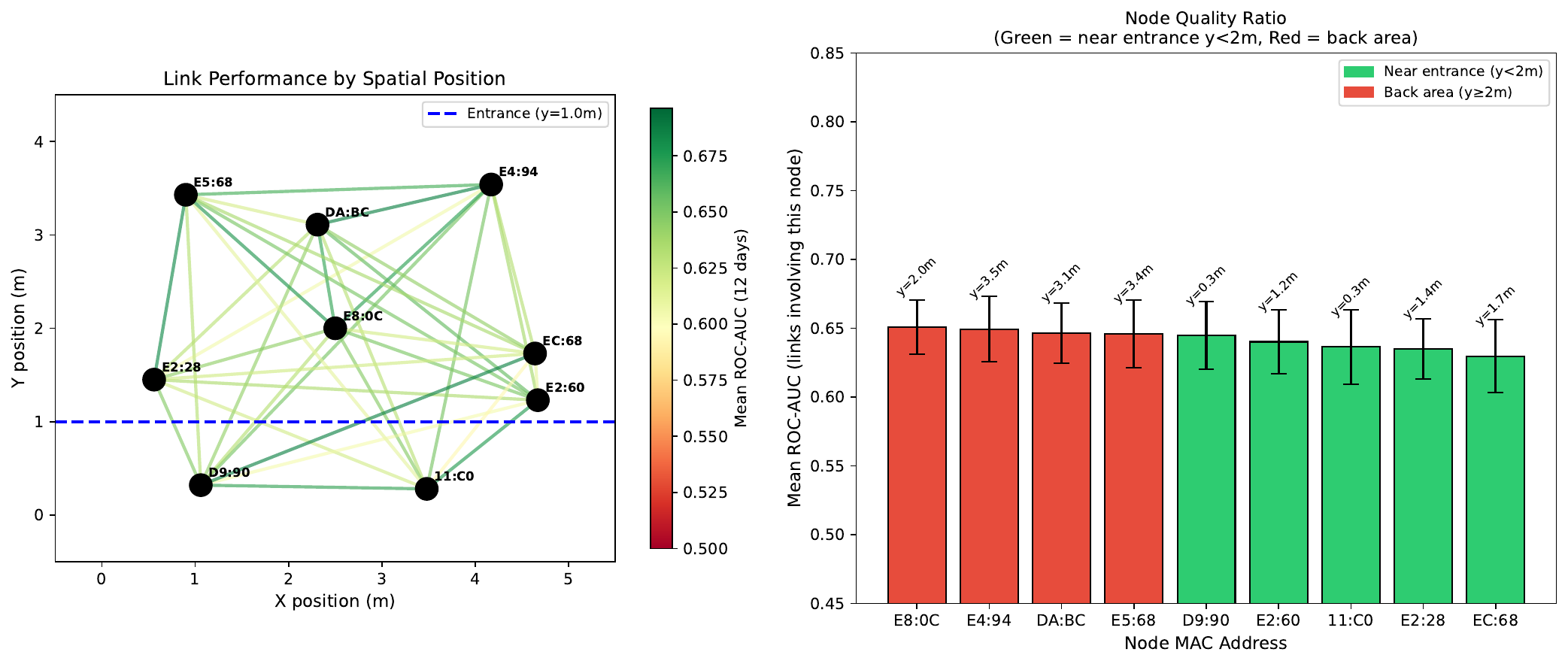}

\vspace{1mm}

\includegraphics[width=\columnwidth]{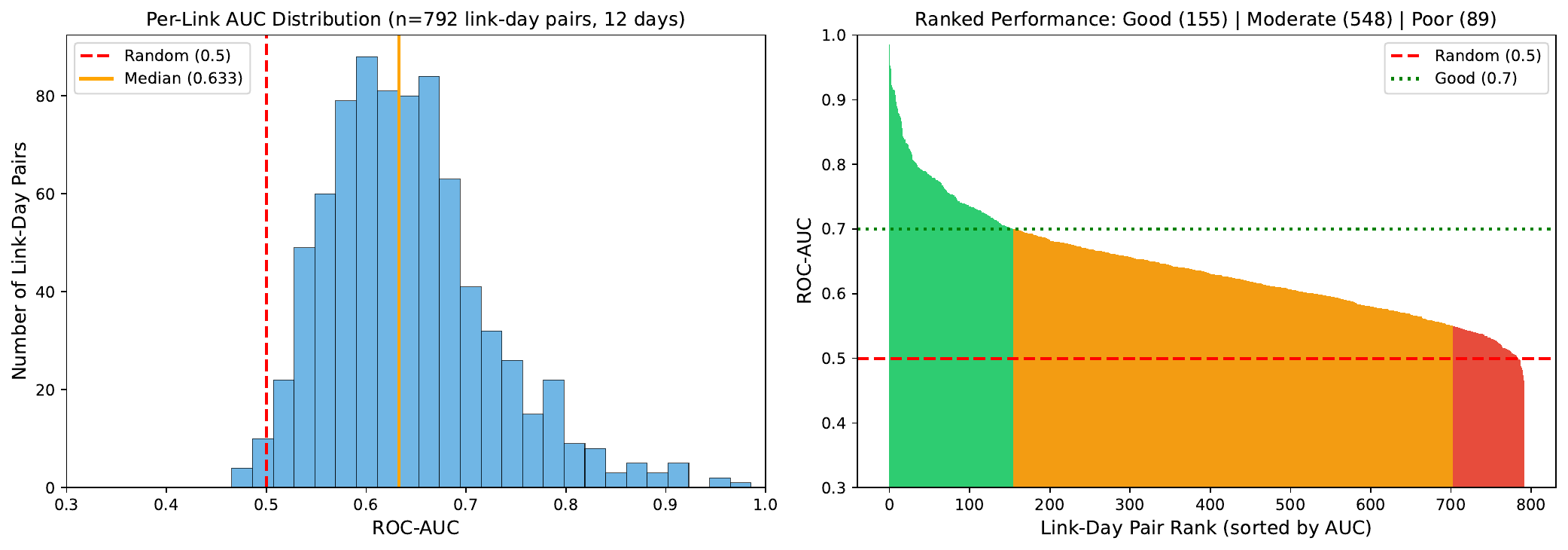}

\caption{Per-link AUC analysis across 12 days (792 link-day pairs). Top: spatial network visualization showing link performance by position (left) and mean ROC-AUC per node with error bars (right); entrance nodes (y$<$2m) and back-area nodes show similar mean AUC ($\sim$0.64). Bottom: distribution showing median AUC of 0.633, with 20\% of link-day pairs achieving AUC$>$0.7.} 
\label{fig:link_quality}
\end{figure}

Figure~\ref{fig:link_quality} reveals the mechanism behind the dilution effect across our full 12-day deployment (792 link-day pairs). The bottom panel shows per-link AUC distribution: median 0.633, with 20\% of link-day pairs exceeding 0.7 and 11\% falling below chance. The 12-day analysis shows that link quality is more variable than short-term evaluations indicate. However, concatenating links degrades performance as observed in the top panel. Link quality shows spatial structure, though with consistent mean AUC across entrance nodes (0.637) and back-area nodes (0.648). Node placement matters, even as individual link rankings shift day-to-day, and in our specific case of a residential deployment, entrance nodes crossed more activity regions.

\begin{findingbox}{Finding 3: Attention cannot rescue multi-link fusion}
Soft attention achieved AUC 0.460$\pm$0.070, \textbf{below random chance}. Hard attention variants performed at exactly random (AUC$\approx$0.50). The attention network cannot determine which links are informative because the per-sample signal is fundamentally weak.
\end{findingbox}

A natural question arises in this scenario: can learned attention mechanisms overcome the dilution effect by automatically weighting informative links?For example, the attention layer can learn that the entrance‑corridor connection is important while the corner‑to‑corner connection is not. Thus, we implemented attention-based fusion that learns importance scores for each of the 72 links, evaluating both soft attention (weighted sum with softmax normalization) and hard attention (top-$k$ selection for $k \in \{1, 3, 5\}$). 


\begin{figure}[t]
\centering
\definecolor{cbBlue}{RGB}{0,114,178}
\definecolor{cbRed}{RGB}{213,94,0}
\definecolor{cbGray}{RGB}{102,102,102}

\begin{subfigure}[b]{0.48\columnwidth}
\centering
\begin{tikzpicture}
\begin{axis}[
    width=\textwidth,
    height=4.2cm,
    ybar,
    bar width=14pt,
    ylabel={AUC},
    ylabel style={font=\small},
    xlabel={Method},
    xlabel style={font=\small},
    symbolic x coords={Top-1, Soft Att.},
    xtick=data,
    xticklabel style={font=\small},
    ymin=0.38, ymax=0.62,
    ytick={0.40, 0.45, 0.50, 0.55, 0.60},
    yticklabel style={font=\small},
    ymajorgrids=true,
    grid style={dashed, gray!30},
    axis lines=left,
    enlarge x limits=0.5,
    clip=false,
]

\addplot[cbGray, dashed, line width=1pt, forget plot]
    coordinates {(Top-1, 0.5) (Soft Att., 0.5)};

\addplot[fill=cbBlue, draw=cbBlue!70!black, line width=0.4pt,
    error bars/.cd, y dir=both, y explicit,
    error bar style={line width=0.8pt, black},
] coordinates {(Top-1, 0.541) +- (0, 0.043)};

\addplot[fill=cbRed, draw=cbRed!70!black, line width=0.4pt,
    error bars/.cd, y dir=both, y explicit,
    error bar style={line width=0.8pt, black},
] coordinates {(Soft Att., 0.460) +- (0, 0.070)};

\node[above, font=\scriptsize\bfseries] at (axis cs:Top-1, 0.59) {0.54};
\node[above, font=\scriptsize\bfseries] at (axis cs:Soft Att., 0.54) {0.46};

\end{axis}
\end{tikzpicture}
\caption{Performance comparison}
\label{fig:attention_auc}
\end{subfigure}
\hfill
\begin{subfigure}[b]{0.48\columnwidth}
\centering
\begin{tikzpicture}
\begin{axis}[
    width=\textwidth,
    height=4.2cm,
    xlabel={Link rank},
    xlabel style={font=\small},
    ylabel={Weight},
    ylabel style={font=\small},
    xmin=1, xmax=72,
    ymin=0.010, ymax=0.022,
    xtick={1, 24, 48, 72},
    xticklabel style={font=\small},
    ytick={0.012, 0.016, 0.020},
    yticklabel style={font=\small},
    ymajorgrids=true,
    grid style={dashed, gray!30},
    axis lines=left,
    legend style={
        at={(0.97,0.97)},
        anchor=north east,
        font=\scriptsize,
        draw=none,
        fill=white,
        fill opacity=0.9,
    },
]

\addplot[fill=cbBlue, fill opacity=0.4, draw=cbBlue, line width=1pt,
] coordinates {
    (1, 0.0192) (18, 0.0173) (36, 0.0158) (54, 0.0150) (72, 0.0146)
    (72, 0.010) (1, 0.010)
} -- cycle;
\addlegendentry{Learned}

\addplot[cbRed, line width=1.5pt, dashed]
    coordinates {(1, 0.01389) (72, 0.01389)};
\addlegendentry{Uniform}

\end{axis}
\end{tikzpicture}
\caption{Learned weights}
\label{fig:attention_weights}
\end{subfigure}

\caption{Attention-based fusion fails: soft attention performs below chance (0.46 vs 0.50) while learned weights remain nearly uniform (max/min~=~1.3$\times$).}
\label{fig:attention-fails}
\end{figure}
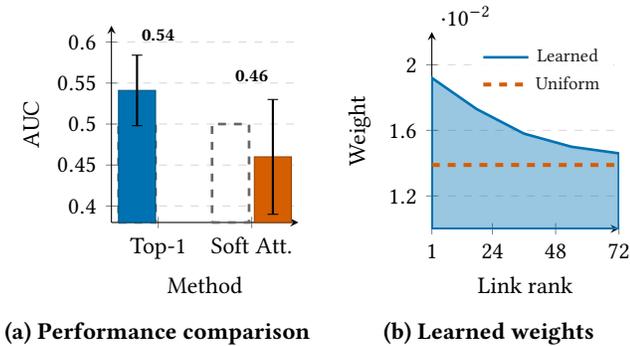

Despite revolutionizing sequence modeling~\cite{vaswani2017attention}, attention mechanisms fail here because they learn link importance from gradient signals during training. When no link provides strong per-sample discrimination, gradients are too noisy to learn meaningful weights. 
Therefore, the model defaults to a uniform or random attention distribution, unable to isolate the sparse "tripwire" links from the 69 uninformative neighbors that contribute only variance without information.

These findings have concrete cost-benefit implications. Our full 9-node mesh costs approximately \$90 in hardware (\$10 per ESP32-C3 node) yet achieves lower accuracy than an optimal 2-node subset costing \$20. Strategic placement thus yields 6$\times$ cost efficiency while simultaneously improving accuracy. 

Beyond cost, link selection enables higher temporal resolution: our TDMA protocol allocates 80~ms per transmitting node, so reducing from 9 nodes to 3 nodes decreases the polling cycle from 720~ms to 240~ms. This increases the effective sampling rate from 1.4~Hz (sub-Nyquist for human walking) to 4.2~Hz. This bandwidth recovery is critical: at 4.2~Hz, the system can track trajectories and detect brief room crossings that the 1.4~Hz configuration completely misses. For example, dead corner nodes (E5:68, 11:C0) had their Fresnel zones partially blocked by furniture and an occupant could remain sedentary for extended periods without triggering them.

A final observation concerns \textbf{temporal stability}. Over our 12-day deployment, the identity of the optimal link was not constant: the top-performing link varied across 4 different physical links, with the ranking shifting approximately every 3 days. This temporal instability likely reflects slow environmental drifts, such as seasonal changes in multipath (\textit{e.g.,} foliage), and, most importantly, variations in occupant behavior between weekdays and weekends. However, while the \textit{identity} of the optimal link is temporally unstable, the \textit{effect} of link selection is robust: in all 10 test folds, Top-1 outperformed All-72. This suggests a practical deployment strategy: periodic recalibration (perhaps weekly) to identify the current best link, rather than one-time optimization during installation.

To enable replication and extension of these findings, we release the complete experimental infrastructure: 312 hours of raw CSI data across all 72 links with synchronized camera-based ground truth labels, the ESP32-C3 firmware implementing our TDMA protocol and binary CSI extraction, and Python analysis scripts reproducing all figures and statistical tests. This dataset addresses the gap between controlled laboratory conditions and real-world deployment~\cite{Turetta2023a, Wang2023Enabling}.

\section{Discussion}
\label{sec:discussion}

We discuss our findings through the lens of Fresnel zone physics and describe implications for WiFi sensing practice.

\noindent\textbf{The dilution effect.} 
The counter-intuitive finding that adding more nodes/links degraded performance comes from Fresnel zone geometry~\cite{fresnel_sensing, Wang2022Placement}. Human presence perturbs CSI only within the ellipsoidal zone between transmitter and receiver, where the first Fresnel zone radius at distance $d$ from endpoints is approximately $r_1 = \sqrt{\lambda d_1 d_2 / (d_1 + d_2)}$. For a 4-meter link at 2.4~GHz ($\lambda = 12.5$~cm), the maximum radius at the midpoint is roughly 50~cm. A human standing 1 meter from this zone produces negligible CSI perturbation because reflected signals arrive out of phase. Thus, adding nodes/links increases the feature space dimensionality without strictly increasing the signal. 

\begin{finding}
Before investing in algorithmic complexity, verify that links intersect activity regions. A simple site survey would identify informative links more effectively than model optimization.
\end{finding}

%
Naive fusion treats all links equally. When concatenating 72 links into a feature vector (72 links $\times$ 11 features = 792 dimensions), the classifier faces a high-dimensional space where discriminative signal is distributed sparsely and inconsistently across links. Standard regularization techniques (L2 penalty, dropout) prevent extreme weight assignments to avoid overfitting, but this same mechanism prevents the model from dynamically selecting the currently-informative links. The classifier cannot learn that link $i$ is informative at time $t$ but link $j$ is informative at time $t+1$. This explains why all five classifiers performed similarly on All-72 (AUC 0.489--0.501): the bottleneck is information-theoretic, not computational.

\noindent\textbf{Resource allocation under sub-nyquist constraints.}
The 2.4~GHz ISM band imposes a fixed ``sensing budget'' that must be allocated wisely. In our backend-orchestrated TDMA protocol, each node requires an 80~ms transmission slot to avoid collisions without GPS synchronization. With 9 nodes, the full mesh cycle requires $9 \times 80 = 720$~ms, yielding a 1.4~Hz sampling rate. Human walking produces limb oscillations at 1--3~Hz~\cite{wifihar2022}, placing our sampling rate at or below the Nyquist rate frequency and risking temporal aliasing that manifests as missed detections during rapid movement.

Reducing from 9 nodes to 3 nodes decreases cycle time to 240~ms, increasing effective sampling rate to 4.2~Hz and providing comfortable headroom above the Nyquist rate limit. At 4.2~Hz, the system can detect brief room crossings that the 1.4~Hz configuration misses. Each additional node in the mesh consumed airtime while contributing to the high-dimensional feature space that overwhelms classifiers. 

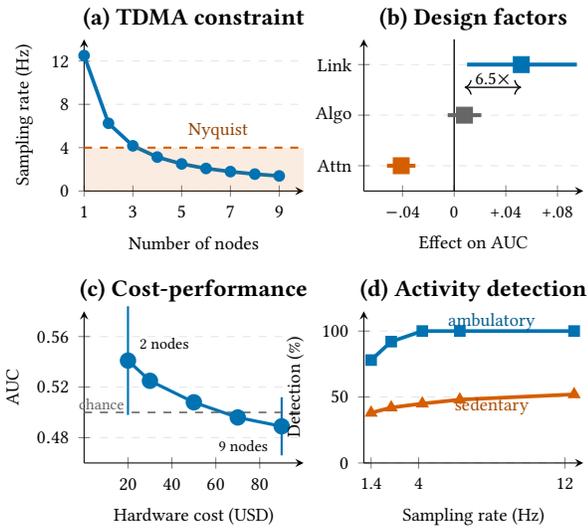
\begin{figure}[t]
\centering
\hspace{-4mm}%
\definecolor{cbBlue}{RGB}{0,114,178}
\definecolor{cbRed}{RGB}{213,94,0}
\definecolor{cbGray}{RGB}{102,102,102}
\definecolor{cbGreen}{RGB}{0,158,115}

\begin{tikzpicture}
\begin{groupplot}[
    group style={
        group size=2 by 2,
        horizontal sep=0.8cm,
        vertical sep=1.6cm,
    },
    width=4.5cm,
    height=3.6cm,
    ymajorgrids=true,
    grid style={dashed, gray!20},
    axis lines=left,
    every axis title/.style={at={(0.5,1)}, anchor=south, font=\small\bfseries},
    every axis label/.append style={font=\scriptsize},
    every tick label/.append style={font=\scriptsize},
]

\nextgroupplot[
    title={(a) TDMA constraint},
    xlabel={Number of nodes},
    ylabel={Sampling rate (Hz)},
    xmin=1, xmax=10,
    ymin=0, ymax=14,
    xtick={1,3,5,7,9},
    ytick={0,4,8,12},
]
\addplot[cbBlue, line width=1.4pt, mark=*, mark size=1.5pt]
    coordinates {(1,12.5) (2,6.25) (3,4.17) (4,3.13) (5,2.5) (6,2.08) (7,1.79) (8,1.56) (9,1.39)};
\addplot[fill=cbRed, fill opacity=0.12, draw=none]
    coordinates {(1,0) (1,4) (10,4) (10,0)} -- cycle;
\addplot[cbRed, line width=0.8pt, dashed] coordinates {(1,4) (10,4)};
\node[font=\scriptsize, cbRed!70!black] at (axis cs:6.5,5.5) {Nyquist};

\nextgroupplot[
    title={(b) Design factors},
    xlabel={Effect on AUC},
    xmin=-0.07, xmax=0.10,
    ymin=0.5, ymax=3.5,
    xtick={-0.04, 0, 0.04, 0.08},
    xticklabels={$-$.04, 0, +.04, +.08},
    ytick={1, 2, 3},
    yticklabels={},
    scaled x ticks=false,
    clip=false,
]
\addplot[black, line width=0.5pt] coordinates {(0,0.5) (0,3.5)};
\addplot[cbBlue, only marks, mark=square*, mark size=3pt] coordinates {(0.052, 3)};
\draw[cbBlue, line width=1.5pt] (axis cs:0.01, 3) -- (axis cs:0.095, 3);
\node[font=\scriptsize, anchor=east] at (axis cs:-0.072, 3) {Link};
\addplot[cbGray, only marks, mark=square*, mark size=3pt] coordinates {(0.008, 2)};
\draw[cbGray, line width=1.5pt] (axis cs:-0.005, 2) -- (axis cs:0.021, 2);
\node[font=\scriptsize, anchor=east] at (axis cs:-0.072, 2) {Algo};
\addplot[cbRed, only marks, mark=square*, mark size=3pt] coordinates {(-0.041, 1)};
\draw[cbRed, line width=1.5pt] (axis cs:-0.052, 1) -- (axis cs:-0.030, 1);
\node[font=\scriptsize, anchor=east] at (axis cs:-0.072, 1) {Attn};
\draw[<->, line width=0.5pt] (axis cs:0.008, 2.55) -- (axis cs:0.052, 2.55);
\node[font=\scriptsize] at (axis cs:0.03, 2.72) {6.5$\times$};

\nextgroupplot[
    title={(c) Cost-performance},
    xlabel={Hardware cost (USD)},
    ylabel={AUC},
    xmin=0, xmax=100,
    ymin=0.46, ymax=0.58,
    xtick={20,40,60,80},
    ytick={0.48, 0.52, 0.56},
    clip=false,
]
\addplot[cbBlue, line width=1.2pt, mark=*, mark size=2.5pt]
    coordinates {(20, 0.541) (30, 0.525) (50, 0.508) (70, 0.496) (90, 0.489)};
\draw[cbBlue, line width=0.8pt] (axis cs:20,0.498) -- (axis cs:20,0.584);
\draw[cbBlue, line width=0.8pt] (axis cs:90,0.466) -- (axis cs:90,0.512);
\addplot[cbGray, dashed, line width=0.6pt] coordinates {(0,0.5) (100,0.5)};
\node[font=\tiny, cbGray] at (axis cs:8,0.507) {chance};
\node[font=\tiny, anchor=south west] at (axis cs:21,0.543) {2 nodes};
\node[font=\tiny, anchor=north east] at (axis cs:88,0.485) {9 nodes};

\nextgroupplot[
    title={(d) Activity detection},
    xlabel={Sampling rate (Hz)},
    ylabel={Detection (\%)},
    xmin=1, xmax=13,
    ymin=0, ymax=115,
    xtick={1.4, 4, 12},
    xticklabels={1.4, 4, 12},
    ytick={0, 50, 100},
]
\addplot[cbBlue, line width=1.4pt, mark=square*, mark size=1.5pt]
    coordinates {(1.4, 78) (2.5, 92) (4.2, 100) (6.25, 100) (12.5, 100)};
\addplot[cbRed, line width=1.4pt, mark=triangle*, mark size=1.5pt]
    coordinates {(1.4, 38) (2.5, 42) (4.2, 45) (6.25, 48) (12.5, 52)};
\node[font=\scriptsize, cbBlue!80!black] at (axis cs:8,107) {ambulatory};
\node[font=\scriptsize, cbRed!80!black] at (axis cs:8,43) {sedentary};

\end{groupplot}
\end{tikzpicture}

\caption{Design trade-offs: (a)~TDMA limits sampling rate as nodes increase; (b)~link placement dominates algorithm choice (6.5$\times$); (c)~sparse deployment outperforms at lower cost; (d)~ambulatory detection saturates above Nyquist.}
\label{fig:discussion-tradeoffs}
\end{figure}

This resource trade-off differs fundamentally from camera-based sensing, where adding pixels never reduces information. In WiFi sensing, the shared medium creates a zero-sum game: spectrum allocated to uninformative links is spectrum denied to informative ones. This constraint is universal: the 2.4~GHz ISM band is congested worldwide, and the physics of Fresnel zones apply regardless of environment. While our study focuses on residential deployment, the principle generalizes to offices, retail, and industrial settings. Figure~\ref{fig:discussion-tradeoffs}a illustrates this constraint: sampling rate drops below the Nyquist threshold at 3 nodes.


\noindent\textbf{Why selection method matters less than expected.} Optimized link selection (Top-1 by SNR) provided no significant advantage over random single-link selection ($p$=0.35). This surprising result deserves careful interpretation (Figure~\ref{fig:discussion-tradeoffs}b shows that link placement dominates algorithm choice by 6.5$\times$). If the benefit of link selection came from identifying the single best link, we would expect optimized selection to substantially outperform random selection. Instead, the near-equivalent performance suggests that benefits came from \textit{avoiding} multi-link fusion rather than \textit{identifying} the best link.

The mechanism is straightforward: even a randomly chosen link avoids the dilution effect that afflicts the 72-link configuration. With most links performing above chance in isolation (median AUC 0.633, with 89\% of link-day pairs exceeding random), a random selection has reasonable odds of picking a useful link while avoiding the high-dimensional feature space that overwhelms multi-link classifiers. The practical implication is liberating: sophisticated link selection algorithms are unnecessary. Simply avoiding fusion provides most of the benefit, and any single link is likely to outperform naive multi-link aggregation. 

\noindent\textbf{Why absolute performance is not great.} The gap between the AUC ($\sim$0.54) and $>$95\% accuracy reported in controlled studies~\cite{Yang2023SenseFi} is noteworthy. We attribute this gap to three factors that distinguish naturalistic deployments from laboratory conditions.

First, \textit{activity composition} differs fundamentally. Laboratory datasets typically feature continuous walking, hand gestures, or exercise activities that produce strong, periodic CSI perturbations. Our naturalistic dataset includes 51 hours of stationary occupied periods (reading, working at a desk) where motionless humans produce CSI perturbations only during infrequent postural adjustments. At our 1.4~Hz sampling rate given by the ESP32-C3, a stationary human is electromagnetically indistinguishable from an empty room.

Second, our \textit{convenience-based node placement} created Fresnel zone geometries that laboratory studies avoid. Researchers optimizing for accuracy would place transmitter-receiver pairs to span activity corridors; we deliberately included corner placements and furniture-occluded positions typical of consumer IoT deployments. 
While individual links performed above chance (median AUC 0.633 across the 12-day deployment), the high-dimensional concatenation degraded overall accuracy. 

Third, our 12-day longitudinal protocol introduced \textit{temporal variability} absent from single-session studies. Environmental drift, weekday versus weekend activity patterns, and gradual changes in furniture arrangement created distribution shift between training and test data. The forward-chaining cross-validation ensures we never train on future data, exposing the model to realistic generalization challenges that same-day train/test splits would obscure.

\noindent\textbf{Implications for system design.} Our findings suggest a shift in WiFi sensing system design philosophy from maximizing coverage to optimizing information density. We propose three concrete design guidelines.

\textit{Guideline 1: Fresnel zone survey before deployment.} Before purchasing hardware, sketch the target environment and identify activity corridors: doorways, hallways, paths between furniture. Links whose Fresnel zones intersect these corridors will carry occupancy-correlated signal; links traversing static space (corners, ceiling regions, furniture-occluded areas) will not. A simple geometric analysis using the Fresnel radius formula ($r_1 \approx 0.5\sqrt{d}$ meters for 2.4~GHz, where $d$ is link distance) can predict which node pairs will form informative links.

\textit{Guideline 2: Minimal viable mesh.} Deploy the smallest node set that creates ``tripwire'' links across mandatory traversal paths. In residential settings, entry points (doors, stairs) are natural choke-points where all movement must pass. Two nodes flanking a doorway create a single high-quality link; adding six more nodes in corners creates 70 additional links that, while individually reasonable, collectively degrade performance through the dilution effect. 
The cost-performance ratio strongly favors sparse, strategic placement: our 2-node optimal subset at \$20 outperformed the 9-node mesh at \$90 (Figure~\ref{fig:discussion-tradeoffs}c).

\textit{Guideline 3: Trade density for temporal resolution.} Each additional node in a TDMA mesh extends the polling cycle, reducing sampling rate. A 9-node mesh at 1.4~Hz samples below Nyquist rate for typical walking speeds; a 3-node mesh at 4.2~Hz provides comfortable headroom (Figure~\ref{fig:discussion-tradeoffs}d shows walking (ambulatory) detection saturates above Nyquist rate while sedentary remains limited). For applications requiring rapid response (fall detection, intrusion alerts), the bandwidth recovered from sparse deployment may be more valuable than the marginal spatial coverage from additional nodes.


\section{Related Work}
\label{sec:related}

We position our work within five research areas, outlining the gap our study addresses.

\noindent\textbf{Foundational WiFi sensing.} WiFi sensing originated with WiSee~\cite{Pu2013WiSee} and WiTrack~\cite{Adib2014WiTrack}. The shift to commodity hardware via the Linux CSI Tool~\cite{csi_fundamentals} and ESP32~\cite{esp32csi} enabled dense deployments like our 9-node mesh.

\noindent\textbf{Deep learning approaches.} SenseFi~\cite{Yang2023SenseFi} benchmarks CNNs, LSTMs, and Transformers achieving $>$95\% accuracy. Domain adaptation~\cite{Sheng2024MetaFormer, hou2024rfboost, Zhang2023SIDA} and gesture recognition~\cite{wigest2015} assume adequate signal quality; they optimize extraction, not collection. Our ablation shows algorithm choice contributes only 0.1\% when link placement is sub-optimal.

\noindent\textbf{Multi-link fusion.} Several systems exploit multiple sensing links~\cite{yan2022joint, lu2022wi, realWorldWiFi2020}, sharing a common assumption: more links provide complementary information. \textit{Critically, none evaluate whether adding low-quality links degrades ensemble accuracy.} Our results challenge this assumption.

\noindent\textbf{Node placement.} Wang et al.~\cite{Wang2022Placement} demonstrate that single-link performance varies by up to 40\% based on placement through Fresnel zone geometry. We extend this analysis to multi-link topologies, investigating which link subset optimizes accuracy when $N$ nodes form $O(N^2)$ possible links.

\noindent\textbf{Occupancy detection.} ESPectre~\cite{espectre2024} achieves F1=98.2\% for motion detection using a single ESP32. Practical deployment studies~\cite{ruvnet2025wifidensepose, Wang2023Enabling} address real-world challenges. Upon expanding to 72 links, we observed signal degradation, contrasting with near-perfect single-link accuracy.

\begin{table}[t]
\caption{Positioning within WiFi sensing literature}
\label{tab:related_comparison}
\small
\resizebox{\columnwidth}{!}{%
\begin{tabular}{lccp{3.5cm}}
\toprule
\textbf{System} & \textbf{Links} & \textbf{HW} & \textbf{Focus} \\
\midrule
WiSee~\cite{Pu2013WiSee} & 1 & SDR & Gesture recognition \\
Widar3.0~\cite{realWorldWiFi2020} & Multi & NIC & Domain transfer \\
SenseFi~\cite{Yang2023SenseFi} & 1 & NIC & DL benchmark \\
Placement~\cite{Wang2022Placement} & 1 & Commodity & Position effects \\
ESPectre~\cite{espectre2024} & 1 & ESP32 & Motion detection \\
\midrule
\textbf{This work} & \textbf{72$\rightarrow$1} & \textbf{ESP32} & \textbf{Link selection for accuracy} \\
\bottomrule
\end{tabular}%
}
\end{table}

\noindent\textbf{Research gap.} As Table~\ref{tab:related_comparison} summarizes, prior multi-link systems assume density improves performance; prior placement work studies single-link optimization. The intersection between these areas remains unexplored. We bridge this gap by identifying a \textit{dilution effect}: we demonstrate that contrary to the prevailing assumption, adding links can degrade accuracy by overwhelming informative signals with feature noise and reducing sampling rates below the Nyquist limit due to bandwidth contention.


\section{Conclusion}
\label{sec:conclusion}

We tested whether denser WiFi sensing deployments yield better accuracy. After 12 days of data collection and evaluation, our answer is a clear \textbf{no}. A single well-placed link (mean AUC 0.541) consistently outperformed a 72-link mesh (mean AUC 0.489) with large effect size (Cohen's $d$=0.86). Algorithm choice had minimal impact; link placement mattered 2.7$\times$ more. 
The bottleneck is therefore information-theoretic rather than computational: when Fresnel zone geometry fails to capture occupancy information, no degree of algorithmic complexity can recover it.

The practical implication is counterintuitive: even links that perform above chance individually (median AUC 0.633 across the 12-day deployment) degrade accuracy when concatenated, due to the dilution effect in high-dimensional feature spaces. 
Even as sensing nodes become smaller, more capable, and affordable, the RF spectrum remains a strictly finite resource; treating it as infinite by deploying dense, unoptimized meshes is physically unsustainable. 
Dense deployments may perform worse than a carefully placed pair of nodes at one-quarter the cost. For practitioners, the most effective intervention may be a floor plan and tape measure rather than a sophisticated model. We release our 312-hour dataset and analysis code to enable validation across diverse environments. 


\balance

\bibliography{references}

\end{document}